\documentclass[10pt,letterpaper]{article}

\usepackage{opex3}
\usepackage[]{hyperref}
\usepackage{amsmath}
\usepackage{epstopdf}
\usepackage{epsfig}
\usepackage{mathbbol}

\begin{document}

\title{Entangled Bessel beams}

\author{Melanie~McLaren,$^{1,2}$ Megan~Agnew,$^{3}$ Jonathan~Leach,$^{3}$ Filippus~S.~Roux,$^1$  Miles~J.~Padgett$^{4}$ and Andrew~Forbes$^{1,2*}$}
\address{$^1$CSIR National Laser Centre, P.O. Box 395, Pretoria 0001, South Africa \\
$^2$Laser Research Institute, University of Stellenbosch, Stellenbosch 7602, South Africa \\
$^3$Department of Physics, University of Ottawa, 150 Louis Pasteur, Ottawa, Ontario, K1N 6N5 Canada \\
$^4$Department of Physics and Astronomy, SUPA, University of Glasgow, Glasgow, UK \\
$^*$Corresponding author: aforbes1@csir.co.za}

\begin{abstract}
Orbital angular momentum (OAM) entanglement is investigated in the Bessel-Gauss (BG) basis. Having a readily adjustable radial scale, BG modes provide a more favourable basis for OAM entanglement over Laguerre-Gaussian (LG) modes. The OAM bandwidth in terms of BG modes can be increased by selection of particular radial modes and leads to a flattening of the spectrum. The flattening of the spectrum allows for higher entanglement. We demonstrate increased entanglement in terms of BG modes by performing a Bell-type experiment and violating the appropriate Clauser Horne Shimony Holt (CHSH) inequality. In addition, we reconstruct the quantum state of BG modes entangled in high-dimensions. 
\end{abstract}

\ocis{(270.0270) Quantum optics, (230.6120) Spatial light modulators}


\section{Introduction}

The use of the orbital angular momentum (OAM) eigenstates of photons in quantum information science became attractive after it was shown, theoretically \cite{Arnaut-2000, Franke-2002} and experimentally \cite{Mair-2001}, that OAM is conserved during spontaneous parametric down-conversion (SPDC) --- the azimuthal indices of a pair of down-converted photons add up to that of the pump beam. As a result, the pair of down-converted photons are naturally entangled in terms of their azimuthal indices or OAM. Entanglement is a desirable property in quantum information applications, where it has been used in quantum ghost imaging \cite{Pittman-1995}, quantum cryptography \cite{Ekert-1991,Gisin-2002} and in quantum computing algorithms \cite{Nielsen-2000}. Any transverse spatial modal basis defines an infinite-dimensional Hilbert space, allowing more information per photon to be used in quantum information applications.

Although the Laguerre-Gaussian (LG) modes are currently the popular choice for theoretical analyses of quantum information systems based on OAM, the modal basis used in quantum information experiments are seldom pure LG modes; see Ref.~\cite{Salakhutdinov-2012}. The optical manipulation and detection of LG modes requires control of the radial index, which governs the radial shape of each LG mode's intensity profile. We shall refer to the modes that are detected without this radial control as azimuthal modes. The azimuthal index is not associated with a unique quantum state, but with a subspace of the total Hilbert space. If the radial part of the mode changes, a photon propagating through an optical system may suffer a loss of its quantum entanglement, even if the azimuthal index is unaffected. As the complexity of quantum information systems increases, the requirements for the fidelity of the quantum states can be expected to become more demanding. Although the connection between the spatial modal profile of an optical beam and OAM was initially made with specific reference to LG beams \cite{Allen-1992}, the same property applies to any optical beam with a rotationally symmetric intensity profile. As a result, Bessel beams \cite{Durnin-1987, Durnin2-1987} and Bessel-Gaussian (BG) beams \cite{Gori-1987} also have quantized amounts of OAM associated with them. Each photon in such a beam also carries an amount of OAM equal to $\ell\hbar$, where $\ell$ is the azimuthal index of the mode.

In this paper, we investigate the use of BG modes \cite{Gori-1987} as a basis for photonic quantum information experiments. The radial index associated with LG modes is replaced by a continuous scaling parameter for the radial part of the BG modes. The relative ease and accuracy with which one can manipulate the azimuthal and radial dependencies of BG modes make them a favourable choice for high-fidelity quantum information systems. We compare the high-dimensional density matrices measured in the BG basis to those measured in the azimuthal basis. Our results indicate that under the correct experimental conditions, the generation and measurement of high-dimensional entanglement in the BG basis is advantageous compared to the azimuthal basis. Our results indicate a possible advantage in using BG modes over azimuthal modes for both generating and measuring OAM entanglement. 

\section{\label{bgspek}Bessel-Gauss Modes}

The electric field of a BG mode is given by 
\begin{eqnarray}
E^{\rm BG}_{\ell}(r,\phi,z) & = & \sqrt{2\over\pi} {\rm J}_{\ell} \left( z_R k_r r \over z_R-iz \right) \exp(i\ell\phi-ik_z z) \nonumber \\
& & \times \exp \left( {i k_r^2 z \omega_0^2 - 2 k r^2 \over 4 (z_R-iz)} \right)'
\label{bgm}
\end{eqnarray}
where $\ell$ is the azimuthal (mode) index (a signed integer); ${\rm J}_{\ell}(\cdot)$ is the Bessel function of the first kind; $k_{r}$ and $k_{z}$ are the transverse and longitudinal wave numbers respectively, obeying $k_{r}^{2} + k_{z}^{2} = k^{2}$. The initial radius of the Gaussian profile is $\omega_0$ and the Rayleigh range is $z_R=\pi\omega_0^2/\lambda$. It is clear from Eq. (\ref{bgm}) that the radial component of the mode can be scaled by altering $k_{r}$.

The quantum state for the photon pairs produced in SPDC can be written in terms of BG modes,
\begin{equation}
\left|\Psi\right\rangle = \sum_{\ell}\!\! \int\!\!\int a_{\ell}(k_{r1},k_{r2}) \left|\ell,k_{r1}\right\rangle_{s} \left|-\ell,k_{r2}\right\rangle_{i} {\rm d} k_{r1} {\rm d} k_{r2},
\label{equ:biphoton}
\end{equation}
where $|a_{\ell}(k_{r1},k_{r2})|^{2}$ denotes the probability amplitude for measuring a signal photon in state $|\ell,k_{r1}\rangle_{s}$ and an idler photon in state $|-\ell,k_{r2}\rangle_{i}$. The radial scaling parameters $k_{r1}$ and $k_{r2}$ in Eq.~(\ref{equ:biphoton}) can be adjusted in a continuous manner to optimize the bandwidth of the OAM spectrum. 

For particular BG modes specified by $k_{r}$, the two-photon state for $d$ dimensions can be written as
\begin{equation}
\left|\Psi\right\rangle = \sum\limits_{\ell} \left|\ell\right\rangle_{s} 
\left|-\ell\right\rangle_{i} ,
\label{Biqudit}
\end{equation}
where $\ell$ ranges over $d$ different values. The sets of $\ell$-values for different dimensions $d$ are chosen as in Ref.~\cite{Agnew-2011}. 

The density matrix of a two-photon qu\textit{d}it quantum state can be reconstructed in terms of a Bloch representation,
\begin{equation}
\rho = \sum\limits_{m,n=0}^{d^2-1} b_{m,n} \tau_{m} \otimes \tau_{n},
\label{eqn:GellMann}
\end{equation}
where $b_{m,n}$ are complex coefficients with $b_{0,0} = 1/d^{2}$ for normalisation; $\tau_p$ are the $d$-dimensional generalised Gell-Mann matrices for $p=1\ ...\ (d^2-1)$; $\tau_0$ is the $d$-dimensional identity matrix. These coefficients are determined by performing a tomographically complete set of measurements. An example of an over-complete set consists of the BG states $| \ell, k_r \rangle$ together with the superpositions of two BG states 
\begin{equation}
| \alpha \rangle = \frac{1}{\sqrt{2}} \left[ |\ell_{1},k_{r}\rangle + \exp(i\alpha) |\ell_{2},k_{r} \rangle \right] .
\label{eqn:super}
\end{equation}
Here, $\alpha$ is the phase between the modes of the superposition states and $k_{r}$ is chosen as a fixed value for all basis states. The quantum state in Eq.~(\ref{equ:biphoton}) consists of two qu\textit{d}its, which results in a $d^{2} \times d^{2}$ density matrix and thus requires at least $d^{4} - 1$ measurements to match the number of matrix elements \cite{James-2001}. However, if an over-complete set of measurements is chosen, additional information is available to compensate for measurement errors due to natural photon number fluctuations. An uncertainty of $5\%$ was observed for the measured coincidence counts. 

\section{Experimental Setup}
\label{sec:setup}

\begin{figure}[htbp]
\centerline{\includegraphics[scale=0.45]{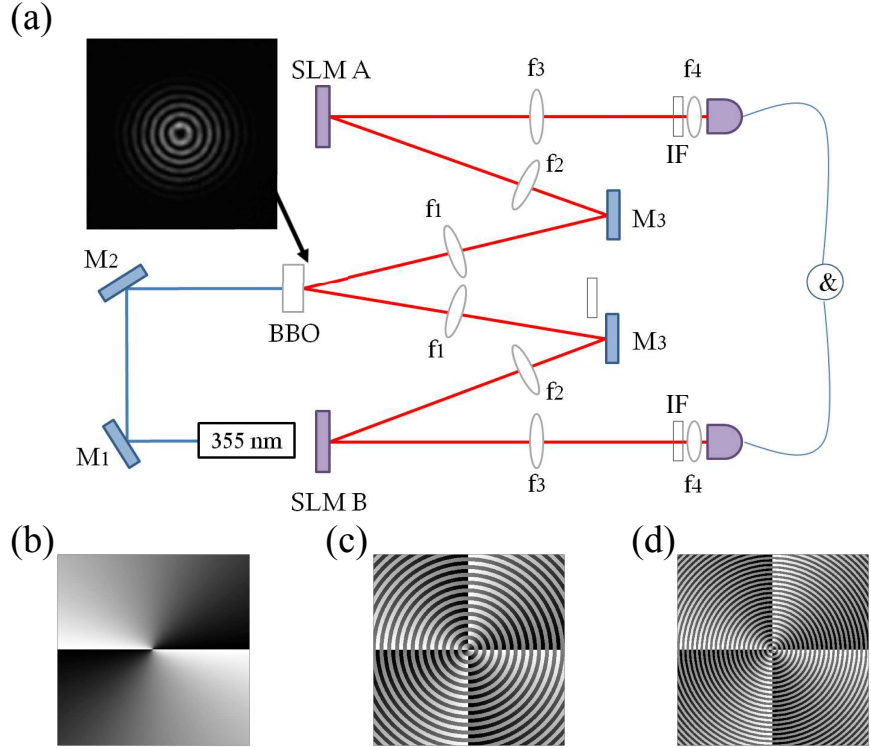}}
\caption{(a) Experimental setup used to detect the OAM eigenstate after SPDC. The plane of the crystal was imaged onto two separate SLMs using lenses, $\rm f_{1} = 400$ mm and $\rm f_{2} = 750$ mm, where the BG modes were selected. Lenses $ \rm f_{3} = 300$ mm and $\rm f_{4} = 1.5$ mm were used to image the SLM planes through 10 nm bandwidth interference filters (IF) to the inputs of the single-mode fibres. Examples of a phase-only binary bessel hologram with a helical phase of $\ell=2$ for different values of $k_{r}$ are shown in (b) $k_{r} = 0$~rad/mm, (c) $k_{r} = 21$~rad/mm and (d) $k_{r} = 35$~rad/mm. The inset shows a back-projected CCD image of a binary bessel mode with helical phase $\ell = 2$ and radial wavevector $k_{r} = 21$~rad/mm measured at the plane of the crystal.}
\label{fig:setup}
\end{figure}

Our experimental setup is shown in Fig.~\ref{fig:setup}. A mode-locked laser source with a wavelength of 355 nm and an average power of 350 mW was used to pump a 0.5-mm-thick type I BBO crystal to produce non-collinear, degenerate entangled photon pairs via SPDC. The plane of the crystal was imaged onto two separate spatial light modulators (SLMs). The SLMs were used to select the particular pair of BG modes that were to be detected, as explained below. The SLM planes were re-imaged and coupled into single-mode fibers to extract the pure Gaussian modal components. The fibers were connected to avalanche photodiodes, which then registered the photon pairs via a coincidence counter. The single count rates $S_{\ell}^{(s)}$ and $S_{\ell}^{(i)}$ were recorded simultaneously with the coincidence count rates $C_{\ell}$ and were accumulated over a 10 second integration time. 

In order to select BG modes using SLMs, we encode a binary Bessel phase function to approximate the Bessel function, thereby avoiding the need to perform complex amplitude modulation \cite{Arrizon-2003}, which results in the loss of optical power. The binary Bessel function is then combined with a helical phase function to determine the azimuthal index $\ell$ of the detected BG mode. The phase-only hologram used to project the photons into BG modes is given by the transmission function
\begin{equation}
T(r) = {\rm sign}\left\{ {\rm J}_{\ell}(k_{r}r)\right\} \exp(i\ell\phi) ,
\label{eqn:phase}
\end{equation}
where sign$\{\cdot\}$ denotes the sign-function. Spatial filtering introduced in each of the detection arms can be used to remove the binary nature of the transmission function, causing the effective modes that are detected to be the BG modes. Without the binary Bessel function, the transmission function is the same helical phase function that is traditionally used to measure the azimuthal modes. However, each OAM mode with a particular value of $| \ell\rangle$ is a superposition of different radial modes that have the same azimuthal index. The binary Bessel function, which is encoded on the SLMs,  enables us to select one particular radial mode from this superposition. As a result of selecting one radial mode, we anticipate two notable features in the measured coincidence counts. Firstly, the spiral bandwidth of the BG spectrum will change depending on the chosen radial mode. Secondly, as only one radial mode from the superposition is selected, the coincidence count rate will be lower when the Bessel function sign$\{J_\ell\}$ is included compared to when it is not. The shape of the spectrum will also change as a result of the binary Bessel function, because one only measures the part (one-dimensional slice) of the full two-dimensional spectrum that corresponds to a particular value of $k_r$.

Following the method derived in Ref.~\cite{Leach-2009}, we used superpositions of OAM states with $\pm\ell$, in analogy to the polarisation states, to demonstrate the violation of a Bell-type inequality. The superposition states for BG modes given by Eq.~(\ref{eqn:super}) were oriented at angles $\alpha_{A}$ and $\alpha_{B}$ for the signal and idler photons, respectively.

Using the method described in Ref.~\cite{Agnew-2011}, we calculate the density matrix for higher dimensions by minimizing the Chi-square quantity,
\begin{equation}
\chi^{2} = \sum\limits_{i = 1}^{N^{2}}{\left[p_{i}^{(M)} - p_{i}^{(P)} \right]^{2}\over p_{i}^{(P)}},
\label{eqn:chi}
\end{equation}
where $p_{i}^{(M)}$ are the normalized coincident counts, which are interpreted as experimentally measured probabilities, and $p_{i}^{(P)}$ are the predicted probabilities calculated from the guessed density matrix. We construct the density matrix so that it is both Hermitian and positive semi-definite with unitary trace \cite{James-2001}. This is necessary as the effects of statistical variations of the coincidence count rates and experimental inaccuracies commonly produce reconstructed density matrices with negative eigenvalues and a non-unitary trace.

\section{Experimental Results}
\label{sec:results}

\begin{figure}[ht]
\centerline{\includegraphics[scale=0.39]{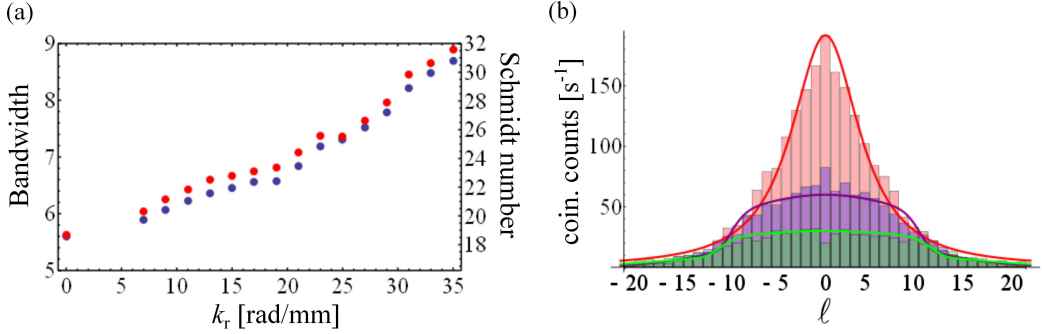}}
\caption{(a) OAM bandwidth (blue) and Schmidt number (red) are shown as a function of $k_{r}$. As the radial wavevector increases from $k_{r} = 0$~rad/mm (azimuthal modes), the OAM bandwidth increases. (b) OAM spectrum for azimuthal modes (red) and BG modes for $k_{r} = 21$~rad/mm (purple) and $k_{r} = 35$~rad/mm (green), in terms of the modal weightings. The maximum coincidence count rate is lower for both BG modes, however a broader spectrum is observed with increasing $k_{r}$.}
\label{fig:bandwidth}
\end{figure}

In the first experiment we vary the radial wave number $k_r$ on each SLM and monitor the coincidence counts for each $\ell$ mode. For $k_r=0$ the phase functions on the SLMs consist of only the helical phase. As a result the $k_r=0$ measurements correspond to the conventional azimuthal modal spectrum. The OAM bandwidth can be quantified as the standard deviation (square root of the second moment) in $\ell$, given by
\begin{equation}
B = \left[ {\sum\limits_{\ell} \ell^{2} C_{\ell} (k_{r}) \over \sum\limits_{\ell} C_{\ell} (k_{r})} \right]^{1/2}
\label{eqn:bandwidth}
\end{equation}
or as the Schmidt number \cite{grobe1994}
\begin{equation}
K = {\left[ \sum\limits_{\ell} C_{\ell}(k_{r}) \right]^{2} \over \sum\limits_{\ell} C_{\ell}^{2} (k_{r})} .
\label{eqn:schnum}
\end{equation}
Both quantities are plotted as a function of $k_{r}$ in Fig.~\ref{fig:bandwidth} (a).

We observe almost a doubling in the OAM bandwidth from $k_{r} = 0$~rad/mm (azimuthal modes) to $k_{r} = 35$~rad/mm (BG modes). This tuneability in the radial component of the field allows for the OAM spectrum to be customised through application of Eq.~(\ref{equ:biphoton}). In Fig.~\ref{fig:bandwidth} (b) one can observe the broadening of the OAM spectrum as the $k_r$ is increased from $k_{r} = 0$~rad/mm (red) to $k_{r} = 21$~rad/mm (purple) to $k_{r} = 35$~rad/mm (green). The maximum coincidence count rate decreases for BG modes, but its spectrum is broader than that of the azimuthal modes. The full-width half-maximum (FWHM) widths of the spectra indicate that the BG modal spectrum at $k_{r} = 35$~rad/mm is almost twice as wide as the azimuthal modal spectrum obtained at $k_{r} = 0$. The wider bandwidth can be explained as an effect of the more selective measurement of the transverse spatial modes, thanks to the specification of the radial definition of the mode (Eq.~\ref{eqn:phase}). As a result one observes a reduction in the central region of the spectrum. Note that the reduction in the central region of the spectrum does not correspond to an attenuation over all radial modes, but rather to a more selective detection of a particular radial mode. The reduction in the central region of the spectrum makes Procrustean filtering in higher dimensional entanglement experiments \cite{Dada-2011} superfluous. Beyond a certain $\ell$-value the BG spectrum coincides with the azimuthal modal spectrum. This can be understood as a consequence of having a large $\ell$-value with a fixed width Gaussian envelope:\ the Gaussian suppression of the function beyond a certain radius causes the definition of the radial dependence of the mode for large $\ell$-values to have little or no effect. Although the spatial resolution of the SLMs would become a limiting factor, larger values of $k_{r}$ are expected to give OAM spectra with larger bandwidths. The wider bandwidth allows more basis states to be incorporated in quantum information systems.

\begin{figure}[htbp]
\centerline{\includegraphics[scale=0.26]{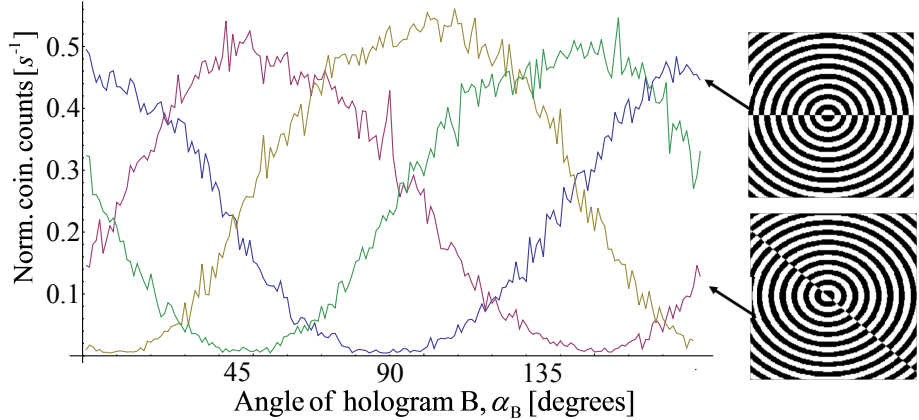}}
\caption{Sinusoidal behaviour of the normalised coincidence counts as a function of the angular position of the holograms, for $\ell = 1$ at positions $\alpha_{A} = 0^{\circ}$ (blue), $45 ^{\circ}$ (pink), $90 ^{\circ}$ (green) and  $135^{\circ}$ (yellow). The insets show the holograms used for $\alpha_{A} = 0^{\circ}$ and $\alpha_{A} = 45^{\circ}$, where the phase varies from 0 (black) to $\pi$ (white).}
\label{fig:bells}
\end{figure}

To verify that the higher-dimensional quantum state produced in the SPDC process is entangled, we use it to test a Bell-type inequality. Clauser, Horne, Shimony and Holt (CHSH) provided a method to experimentally test Bell's theorem, stating that no local hidden-variable theory could simulate the kind of statistical predictions made by quantum mechanics \cite{CHSH-1969}. We fixed $\alpha_{A}$ from Eq.~(\ref{eqn:super}) at four different angles: $\alpha_{A} = 0^{\circ}, 45 ^{\circ}, 90 ^{\circ}, 135^{\circ}$, allowing $\alpha_{B}$ to vary from $0^{\circ}$ to $180^{\circ}$. The Bell parameter, as defined in Ref.~\cite{Leach-2009}, is given by
\begin{equation}
S = E(\alpha_{A}, \alpha_{B}) - E(\alpha_{A}, \alpha_{B}') +E(\alpha_{A}', \alpha_{B}) - E(\alpha_{A}', \alpha_{B}'),
\label{eqn:Sparameter}
\end{equation}
where
\begin{equation}
E(\alpha_{A}, \alpha_{B}) = \frac{C(\alpha_{A}, \alpha_{B})+C(\alpha_{A}+\frac{\pi}{2\ell}, \alpha_{B}+\frac{\pi}{2\ell})-C(\alpha_{A}+\frac{\pi}{2\ell}, \alpha_{B})-C(\alpha_{A}, \alpha_{B}+\frac{\pi}{2\ell})}{C(\alpha_{A}, \alpha_{B})+C(\alpha_{A}+\frac{\pi}{2\ell}, \alpha_{B}+\frac{\pi}{2\ell})+C(\alpha_{A}+\frac{\pi}{2\ell}, \alpha_{B})+C(\alpha_{A}, \alpha_{B}+\frac{\pi}{2\ell})} ,
\label{eqn:Eparameter}
\end{equation}
with $C(\alpha_{A}, \alpha_{B})$ being the coincidence count rate for the particular orientation of each hologram. We see a clear violation of the CHSH inequality for BG modes ($k_{r} = 21$~rad/mm) --- our Bell curves are shown in Fig.~\ref{fig:bells}. For the two-dimensional BG subspace of $\ell = 1$, we find the Bell parameter in the CHSH inequality to be $ S = 2.78 \pm 0.05 > 2$, which violates the inequality by 15 standard deviations.

We also reconstructed high-dimensional two-photon density matrices in the BG basis for $k_{r} = 21$~rad/mm using a full quantum state tomography \cite{Jack-2009}. The density matrices for $d = 2$ and $d = 5$ are shown in Fig. (4). The purity of the state $\rm T \rm r(\rho^{2})$ was calculated as $0.981$, where its closeness to 1 indicates that the observed quantum state is close to a pure state. To quantify the mixture of the measured state we calculate the linear entropy \cite{Bose-2000} ${ S}_{L} = 4/3[1 - {\rm Tr}(\rho^{2})]$. We find the linear entropy of our quantum state to be $ S_{L2} = 0.06$, indicating that our state is close to a pure state (the linear entropy for a pure state is zero, whereas a completely mixed state is charaterised by a linear entropy of one). 

\begin{figure}[htbp]
\centerline{\includegraphics[scale=0.4]{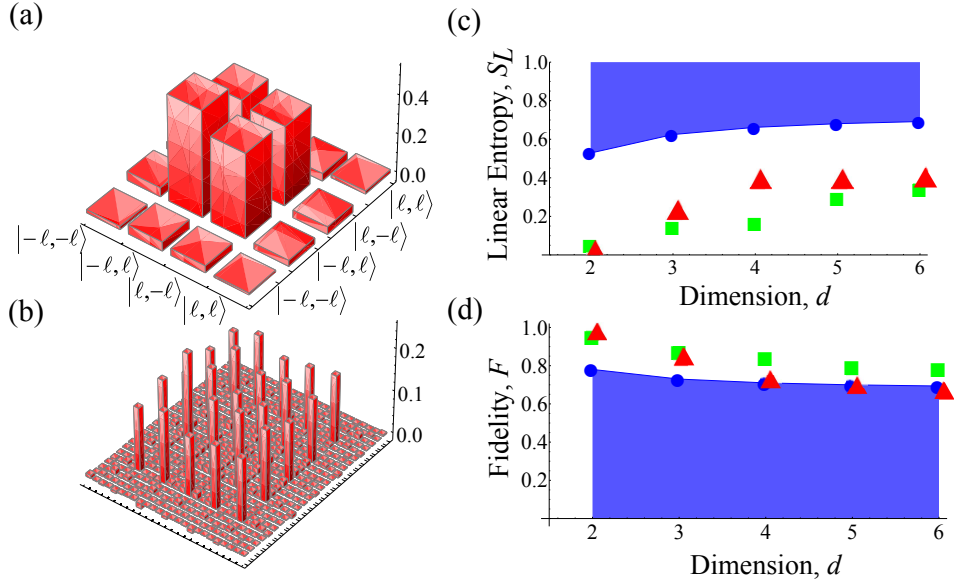}}
\caption{Results from a full quantum state tomoography of a BG mode with $k_{r} = 21$~rad/mm. (a) \& (b) Graphically representation of the real part of the density matrix for dimension, $d = 2$ and $d = 5$, respectively. (c) Linear entropy and (d) fidelity as a function of dimension. The red triangles represent the measured data for the azimuthal modes, the green squares represent the measured data for the BG modes and the blue circles represent the threshold states in Eq.~(\ref{eqn:highBell}).}
\label{fig:highD}
\end{figure}

We also calculate the fidelity
\begin{equation}
F = \left[ {\rm Tr} \left\{ ( \sqrt{\rho_{T}} \rho_{\rm d} \sqrt{\rho_{T}} )^{1/2} \right\} \right]^{2} ,
\label{eqn:fidelity}
\end{equation}
which is a measure of how close our reconstructed state $\rho_{\rm d}$ is to the target state $\rho_{T}$. In our case the target state is the (pure) maximally entangled state. We find a fidelity of $F = 0.96$. These results are indicative of entanglement of our BG modes.
We performed high-dimensional state tomography measurements from $d=2$ to $d=8$ and use the result to compute the fidelity and the linear entropy, as indications of entanglement and purity, respectively. We found that the higher-dimensional quantum states in terms of the BG modes for a particular $k_r$ have higher fidelity than the quantum states in terms of azimuthal modes, while remaining relatively pure, as shown in Fig.~\ref{fig:highD}(c) and (d). These measurements are compared with the threshold states, which are given by
\begin{equation}
\rho_{B} = \rm p_{d}^{min}\left|\psi\right\rangle\left\langle\psi\right| + (1 - \rm p_{d}^{min})\frac{\mathbb{I}}{d^{2}} ,
\label{eqn:highBell}
\end{equation}
which lie on the threshold of the high-dimensional Bell inequality \cite{Collins-2002}. Here $\rm p_{d}^{min}$ is the probability above which the Bell inequality is violated and $\mathbb{I}$ is the identity matrix of dimension $d^{2}$. The fidelity values ranged from $F_{2} = 0.96$ to $F_{6} = 0.79$ for the BG modes, while the values for linear entropy were found to be $S_{L2} = 0.06$ and $S_{L6} = 0.35$. 

\section{Discussion and Conclusion}
\label{sec:concl}
 
We have successfully demonstrated that BG modes are entangled in the high-dimensional OAM degree of freedom. Crucially, the ability to select the radial component allows us to record BG states with a very broad spiral spectrum. The resulting high-dimensionally entangled states are closer to the appropriate maximally entangled state compared to the case where no radial selection is performed. Such control of the experimental conditions results in a higher information capacity per photon pair and allows for more states to be utilized in quantum information processes. The only cost of the increased spiral spectrum of the recorded modes is a lower coincidence count rate, which is due to the required modal selectivity. Due to the particular interference properties of BG modes, entangled modes in this basis may offer increased robustness when propagating through turbulent media. The finite space-bandwidth product of the SLMs limits the achievable spectrum of the BG modes; however, alternative methods may offer further improvements in OAM entanglement with BG modes.

\end{document}